# Electron Energy-Loss Spectroscopy and the Electronic Structure of $ABO_3$ Ferroelectrics: First Principle Calculations

**Sevket Simsek, Amirullah M. Mamedov, Ekmel Ozbay**

*Nanotechnology Research Center (NANOTAM), Bilkent University 06800*

*Bilkent, Ankara Turkey*.

Corresponding author. Email: mamedov@bilkent.edu.tr

# Electron Energy-Loss Spectroscopy and the Electronic Structure of $ABO_3$ Ferroelectrics: First Principle Calculations

The electronic structures of $ABO_3$ ferroelectrics are calculated within the density functional theory, and their evolution is analyzed as the crystal-field symmetry changes from cubic to rhombohedral via tetragonal phases. Electronic structure fingerprints that characterize each phase from their electronic spectra are identified. We carried out electron-energy loss spectroscopy experiments by using synchrotron radiation and compared these results to the theoretical spectra calculated within DFT-LDA. The dominant role of the $BO_6$ octahedra in the formation of the energy spectra of $ABO_3$ compounds was demonstrated. Anomalous behavior of plasmons in ferroelectrics was exhibited by the function representing the characteristic energy loss in the region of phase transition.

**Keywords:** ferroelectrics, energy losses, plasmons

## 1. Introduction

Over the past decades, electron energy loss (EEL) spectroscopy has developed into a major tool for the characterization of nonlinear structures and electronic structure of materials. Most of these studies focus on the ionization edges, where the position of the edges and their intensity provides a quantitative measure of the composition [1]. The fine structure on the edges gives information about unoccupied density states, and hence local bonding. The low-loss energy region of the EEL spectrum (<50 eV) can also provide information about composition and electronic structure, as well as optical properties [1,2], although it has not found as wide application as the energy-loss near-edge structure. In this low-loss region, interband transitions and plasmon losses are observed. Plasmon losses correspond to a collective oscillation of the valence electrons and their energy is related to the density of valence electrons. Consequently, if the valence electron density changes with the composition, the change in the plasmon energy can be used to measure composition. Some examples of plasmon spectra being used to describe different phases and composition are briefly reviewed in [3].

The plasmons hold a unique position among the quasiparticles of solids because of their special features. The spectra of plasmons are described by the functions $-\mathrm{Im}\varepsilon^{-1}$ (volume plasmons) and $-\mathrm{Im}(\varepsilon+1)^{-1}$ (surface plasmons). Experimentally, they are determined by measuring the characteristic electron energy loss (EEL) [4]. In the general case, the loss function has an intricate form because of the superposition of various effects, among them the excitation of plasmons, interband transitions and metastable excitons.

In a complicated problem to extract the functions $–Im\varepsilon^{-1}$ and $–Im(\varepsilon+1)^{-1}$ from EELS (W). For this purpose, various simplifications, approximations, and calibrations are made. The half-width of the principal peak of W typically exceeds 4.0 eV. The intensity of peaks of W critically depends on the orientation of the sample and the energy of the electron beam, their resolution being not finer than 1.0 eV. This severely hampers the determination of the true spectra of plasmons and their energies. For this reason, even for the simplest crystals such as Si and diamond, the experimental data on the EEL spectra are highly contradictory as far as the nature and energies of plasmons of both the types are concerned [5,6]. Of great interest, in this connection, is a calculation procedure in which the experimental data on the reflection spectra and the Kramers-Kronig integral relations are used to determine the plasmon spectra [7].

The plasmon energy can be calculated using free-electron theory as

$$E = \hbar\sqrt{\frac{ne^2}{\varepsilon_0 m}} \tag{1}$$

Where n is a number of valance electrons per unit volume, m is the effective mass of electron. Using the rest mass of electron, $m=m_0$, free-electron theory works reasonably well for many systems. The effect of the underlying ion-core lattice on the plasmon energy can be included by using the effective mass m, rather than $m_0$. However, there is no straightforward approach to calculating m. There are other peaks and features in the low-loss spectrum, in addition to the plasmon peak, associated with interband transitions. Interband transitions are an effect lattice, they shift the plasmon energy and can be included via Drude-Lorentz theory [3,7]. However, in order to include interband transitions, we must know where they are expected to occur, that is, we need to have a good understanding of the electronic structure, a rather circular process if one is trying to use the low-loss spectrum, after the zero-loss peak. It is somewhat surprising that there has not been more effort to improve the calculations of the expected plasmon energy.

The aim of this paper is to apply DFT band structure calculations of the electronic structure, dielectric functions, and spectra of plasmons of both types to a wide variety of some $ABO_3$ (A=Ba, Sr, Li, and K, B=Ti, Nb and Ta) materials in order to explore how well-loss EELS can be predicted.

Our paper is organized as follows. In section 2, we describe the methodology, structure and computational details. In section 3, we describe the computation and theory of EELS. In section 4-5, we illustrate the validity of the formalism by applying methodology and theory (see sections 2 and 3) to $ABO_3$ ferroelectrics. Theoretical analysis of the temperature dependences of the loss spectra near structural and ferroelectric phase transitions we describe in section 6.

## 2. Computational Details

The optical properties of $ABO_3$ were theoretically studied by means of first principles calculations in the framework of density functional theory (DFT) [8] and based on the local density approximation (LDA) [9] as implemented in the ABINIT code [10,11]. The self-consistent norm-conserving pseudopotentials are generated using Troullier-Martiens scheme, which for the $ABO_3$, include the semicore s and p states of A- and B- atoms (see Table 1) [12] which is included in the Perdew-Wang [13] scheme as parameterized by Ceperly and Alder [14]. For the calculations, the wave functions were expanded in plane waves up to a kinetic-energy cutoff of 40 Ha (tetragonal and rhombohedral $KNbO_3$), 38 Ha ($BaTiO_3$, $SrTiO_3$). The level of accuracy for the Kohn-Sham eigenvalues and eigenvectors is required to calculate the response function. The Brillouin zone was sampled using an 8 × 8 × 8 the Monkhorst-Pack [15] mesh of special k points. Rhombohedral positions coordinates of $LiNbO_3$ and $LiTaO_3$ using both experimental value [16-18] were calculated to relate the hexagonal coordinates given in the literature by the transformation [16]. The coordinates of $KNbO_3$ [16-18] and $BaTiO_3$ [16-18] are reported in Table 1. All the calculations of $ABO_3$ have been used with experimental lattice constants and atomic positions. The lattice constants and atomic positions are given in Table 1. The coordinates of the other atoms can easily be obtained by using the symmetry operations of the space groups. These parameters were necessary to obtain converged results in the optical properties.

## 3. Theory

The EEL spectrum can be described in a dielectric formulation [3,19] by

$$\frac{d^2G}{d\Omega\, dE} = \frac{1}{\pi^2 a_0 m_0 v^2 n_a}\left(\frac{1}{\theta^2 + \theta_E^2}\right)\mathrm{Im}\left(-\frac{1}{\varepsilon(q,E)}\right) \tag{2}$$

Where v is the speed of the incident electron, $n_a$ is the number of atoms per unit volume, and $\theta_E$-is the characteristic scattering angle $(\theta_E = E/\gamma m v^2)$. The term $\mathrm{Im}(-\frac{1}{\varepsilon})$ is often called the loss-function. Using eq(2) and a Kramers-Kronig analysis, it possible to obtain the complex dielectric function $\varepsilon = \varepsilon_1 + i\varepsilon_2$ from the low-loss EEL. Optical properties can be calculated from ε. Although loss straightforward than using optical techniques directly, measuring optical properties with EELS offers advantages of better spatial resolution and it extends to higher energies. We can rewrite $\mathrm{Im}(-\frac{1}{\varepsilon})$ as $\frac{\varepsilon_2}{(\varepsilon_1^2 + \varepsilon_2^2)}$ and the plasmon energy (the maximum of the loss-function) is identified as occurrig at the point where $\varepsilon_1$ crosses zero with a positive slope $(d\varepsilon_1/dE > 0)$, and $\varepsilon_2 \ll 1$ with a negative slope $(d\varepsilon_2/dE < 0)$. It follows from eq (2) that, if we want to use an ab initio method to calculate the low-loss EELS we must calculate ε(ω). The dielectric function describes the response of the material to a time-dependent electromagnetic field and the underlying theory is well developed [19].

If the material is anisotropic, the microscopic dielectric function will be a tensor $\varepsilon_{ij}$ and can be calculated [20] as

$$\mathrm{Im}\varepsilon_{ij}(\omega) = \frac{4\pi e^2}{m^2\omega^2} \sum_{c,v} \int dk < c_k \left| P^i \right| v_k >< v_k \left| P^j \right| c_k > \delta(E_{ck} - E_{vk} - \omega) \quad (3)$$

Where $|v_k\rangle$ and $|c_k\rangle$ are the initial and final electron states, respectively, with energies $E_{vk}$ and $E_{ck}$. The $P^i$ are components of the momentum operator, $P = -i\hbar\Delta$ and $\omega = E/\hbar$. Where $v_k$ and $c_k$ are in the same band we have an intraband transition and an interband transition if they arise from different bands.

## 4. Electronic structure of the valence and conduction bands

In this section, we investigate the evolution of the electronic structure of $ABO_3$ in the structural and ferroelectric phase transitions in relation with the evolution of the symmetry of the crystal field. Results from VUV reflectivity experiments are reported in the literature for some $ABO_3$ [21,28-33,36-40]. We furthermore give a comparison of the theoretical band structure of the valence and conduction bands in various phases. Last, we report our results for the theoretical gap, and the gap values determined experimentally.

Since $ABO_3$ ferroelectrics have been vilely studied (see Ref [21-43]) we show in detail only the calculated band structures for $KNbO_3$ in different phases. For all other calculated compounds we presented the main data in Tables 2-3.

The dispersion of the energy of the valence bands for the cubic, tetragonal and rhombohedral phases at the equilibrium lattice parameters are reported in Fig.1-3. The top of valence band is located at R point in cubic and tetragonal phases, and in the Z point in rhombohedral phase. The DOS reported for all phases in Fig. 4-6 show a much higher density of states at and near the top of the valence bands. The valence bandwidth decreases by 0.3 eV between the different crystal fields, from 5.6 eV in the cubic phase to 4.3 eV in the rhombohedral phase. The bandwidth of the tetragonal phase is closer to the width of the cubic phase than to the rhombohedral one.

As we said herein above, the electronic band structure of paraelectric cubic $KNbO_3$ along the symmetry lines of the cubic Brillouin zone is shown Fig. 1. Let us describe the electronic band structure of the $KNbO_3$ in more detail. It is clear that the indirect bad gap appears between the topmost valence band and at the R point and at the bottom of conduction band at the Γ point. The overall profile of our band structure is qualitatively like the band structure obtained by previous studies [24,25]. It is observed that the conduction band minimum goes from the Γ point through Δ to the X point and always remains nearly flat in agreement with previous studies [32,35]. The calculated indirect band gap (R-Γ) is 1.54 eV while the smallest direct band gap ($\Gamma_V$-$\Gamma_C$) is 2.51 eV. These calculated values are smaller than the experimental value of 3.1 eV for the indirect gap [33]. The origin of this discrepancy could be the use of DFT, which generally underestimates the band gap in semiconductors and insulators [8,9]. The band with the lowest energy in Fig. 1, lying between -16.0 eV and -17.0 eV, correspond, to a very large extent, to O 2s states. The nine valence bands are between -5.9 eV and Fermi level (zero) are mainly due to the oxygen O 2p states hybridized with Nb 4d states. These nine valence bands are split into triple and double degenerate levels at the Γ point ($\Gamma_{15}$,$\Gamma_{25}$, $\Gamma_{25}$) separated by energies 1.64 eV ($\Gamma_{15}$-$\Gamma_{25}$), 0.3 eV ($\Gamma_{25}$, $\Gamma_{15}$), and 1.94 eV ($\Gamma_{15}$-$\Gamma_{25}$) due to the crystal field and electrostatic interaction between mainly O 2p and Nb 4d orbitals. In the conduction band, the one three double ($\Gamma_{12}$) degenerate levels represent Nb 4d $t_{2g}$ and Nb $e_g$ orbitals separated by energy 4.2 eV. The topmost valence bands are the oxygen $2p_x$, $2p_y$ states while the lowest valence bands are formed by hybridization of Nb 4d $e_g$ and O $2p_z$ states. In the conduction band region, the first conduction band from about 1.59 eV above the Fermi level to 5.6 eV arises from predominantly Nb $t_{2g}$ states with small O 2p mixing. The bands in the conduction band that are shown in Fig-.1belong to Nb 4d $e_g$ states. Also, some electrons from Nb 4d transform into the valence band and take part in the interaction between Nb and O. This implies that there is hybridization between Nb 4d and O 2p. Conversely, in the conduction band, the DOS (see Fig. 4-6) of Nb d is much higher than that of O 2p. This implies that there are few O p electrons which transform into the conduction band and hybridize with Nb d electrons. The DOS of Nb-(d) and O-(p) thus show that interaction between Nb and O is covalent. On the other hand, the DOS of K 3p shows a peak around 0.5 eV attributed to the bands around this energy in the band structure. The structures lowest in energy between approx.-17.5 eV and 16.0 eV are shown to be of predominantly O 2s character with some mixing of Nb p states. On the whole, our results are in agreement with the LAPW calculation and MF calculation of [34].

The agreement between our DOS and the experimental spectrum is good [21-43]. The dispersion curve of the conduction bands at the theoretical equilibrium lattice parameters are reported in Fig. 1-3. The bottom of the conduction band is found to be at the Γ point for all phases. The first conduction bands are found to have an $e_g$ character at the Γ point in the cubic phase (Fig. 1). At higher energy, conduction bands have $a_{1g}$ or $t_{2g}$ character at the R and Γ points. The gap in the conduction band has a minimum of 1.3 eV at Γ point.

The fundamental gap of the single crystal of $KNbO_3$ has been measured in the cubic and tetragonal phases [28]. More precisely, VUV spectroscopy has been performed in reflectivity in the 1.0-35.0 eV range, and the absorption spectrum has been deduced from a Kramers-Kronig analysis [32]. The band gap obtained from a fit of the low absorption is found to be 2.86 eV. However, EELS experiments a large momentum transfer [33], give a gap value about 2.9 eV. Our fitted value for $E_g$ is 2.45 eV in cubic phase of $KNbO_3$.

We have also calculated the minimum and direct LDA band gap energy at the theoretical lattice parameters (Table 2). Within the accuracy of our calculations, the values of the direct and indirect are within 2.35 eV of each other in all phases except the rhombohedral one, where the minimum gap is 1.5 eV smaller than the first direct gap. Our results are in good agreement with a previous calculation [34] performed at the experimental lattice parameters, with the exception of the rhombohedric phase.

The O 2s dispersion curves are reported in Fig. 1-3. As was the case for the B 3s and 3p lines, the O 2s peak position is lower by 0-2.0 eV. However, in contrast to the B 3s and 3p lines, the oxygen 2s band is very sensitive to the oxygen local coordination. We find that, the O 2s band shows a left shoulder at -2.0 eV with respect to the main peak. Second, we see that the oxygen line shape is identical for the cubic and tetragonal structures. In comparison with the experimental results we note that experimentally [21,33] the O 2s band is found to be centered at -17 eV in the tetragonal phase. In contrast, we find our O 2s doublet located at -16.1 eV and -16.4 eV.

## 5. Electron Energy Loss Spectroscopy

In this section, we first apply the theoretical framework defined in section 3 and report our theoretical EELS spectra. Valance EELS experiments have previously been performed in transmission at a large momentum transfer q ( ref [1] ) and in reflection ( ref [1,28] ). In this work, we have performed very low q and VUV reflectivity [21,28-31 ] experiments and compared them to the theoretical results (see Fig. 7).

In Fig. 8-9 We report (-as an example, because it is well known that optical properties of many $ABO_3$ materials are very close to each other in the energy region up to 30 eV) the dielectric functions (real and imaginary parts and $\text{Im}\varepsilon^{-1}$) for $KNbO_3$ in three phases: cubic, tetragonal, and rhombohedric, calculated in LDA. Three regions can be distinguished. First, we observe that the valence excitation region extends up to 15.0 eV. The form of the structure and the shape of $\varepsilon_2$ for the investigated crystal are determinate by the positions of the critical state density points (see Table). The similarity between the $\varepsilon_2$ for all modifications of $KNbO_3$ in the region up to 14.0 eV points to the substantial role of $NbO_6$ octahedron in the formation of the band structure. This means that the $NbO_6$ octahedron determines the lowest limit of the conduction band and the upper valence band. These bands are similar for many $ABO_3$, since the d-orbitals of the transition metals and the p-orbitals of oxygen, which are joined in each octahedron, make the main contribution to the bands indicated above. The real part of $\varepsilon$ behaves mainly as a classical oscillator. It vanishes (from positive to negative value) around 5.52, 10.01, 11.88, and 13.48 eV, and correspondingly, $\varepsilon_2$ shows maxima of absorption at these frequencies. The real part $\varepsilon$ vanishes (from negative to positive) at 6.40, 11.29, 12.72, and 14.96 eV (not seen in the figure because of the small broadening). The loss function consequently shows peaks at these energies: 6.50 and 15.00 eV. In the absorption spectrum $\varepsilon_2$ the strong absorption below 8.0 eV stems from transitions from the valance band to the $e_g$ states. The absorption band expending beyond 8.5 eV up to 14.0 eV is associated with transitions from the valence band to $t_{2g}$ states in the conduction band. Second, we see that above 15.0 eV, corresponding to the O 2s and Nb 4p excitations, $\varepsilon_1$ also behaves as a classical oscillator: it vanishes (from positive to negative) at 21.30 eV. Peaks are observed in the loss function when $\varepsilon_1$ vanishes (from negative to positive) at 22.11 eV. Third, we remark that the region above 22.0 eV cannot be interpreted in terms of classical oscillators. Above 22.0 eV $\varepsilon_1$ and $\varepsilon_2$ are dominated by linear features, increasing for $\varepsilon_1$ and decreasing for $\varepsilon_2$. The corresponding loss function exhibits a broadened peak at 41.3 eV that we assign not to plasmons but to their forms of collective excitations. The plasmons is defined by a vanishing real part of dielectric function and a minimum of the imaginary part, which is not the case for this peak. Such linear behavior for $\varepsilon_1$ and $\varepsilon_2$ has already been observed in the theoretical EELS of $ABO_3$. At higher energies, however, they drastically modify the triple 4p plasmons, both in line shape and peak position. We also find a small anisotropy for the xx and zz directions. The same situation we can observe in ref. [45] and [46-47] for $ZrO_2$ and $TiO_2$. Some of the $ABO_3$ ferroelectrics that we experimentally investigated had convenient (from the point of view of the apparatus) PT (phase transition) points $T_c$ ($KNbO_3$, and others ferroelectrics). This enabled us to carry out a study of the temperature dependences in the PT region on $KNbO_3$.

The temperature dependences of $-\text{Im}\varepsilon^{-1}$ constructed by us for a fixed value of the ($\hbar\omega$=23.0 eV) demonstrated the anomalous change in $-\text{Im}\varepsilon^{-1}$ in the region of $T_c$ (Fig. 10). It is clear from Fig. 10 that curve exhibited a clear maximum in the loss spectrum of $KNbO_3$ at $T\simeq940$ K. Unfortunately, the complexity of the experiments and large volume of calculation work, as well as the difficulties encountered with the stabilization of temperature in the region $T_c$ prevented us from investigating in detail the anomalous behavior of $-\text{Im}\varepsilon^{-1}$ in $KNbO_3$ in the PT region. We were able to simply report an anomaly in the characteristic loss spectra in the vicinity of $T_c$. Naturally, such anomalous behavior of $-\text{Im}\varepsilon^{-1}$ should be associated with ferroelectricity in these crystals. Similar anomalous behavior of the losses was observed by us for $Gd_2(MoO_4)_3$ [48]. Recently, there have been some reports [49] in which an analogous effect was observed in $BaTiO_3$ and in triglycine sulfate. In the next section, we try to explain the anomalous behavior of characteristic energy loss function on the base of dependence of high-frequency permittivity $\varepsilon(\omega)$ on polarization.

**6. Anomalous of Energy Losses near phase transitions**

Energy losses of fast electron in ferroelectric crystals are described by the following expression common to condensed media

$$W = -\frac{8\pi e^2}{\hbar K^2}\,\text{Im}\frac{1}{\varepsilon(\vec{K},\omega)} \qquad (3)$$

That is simpler than (2) and, where $\hbar K$ - is a pulse of transfer, and the dielectric constant $\varepsilon(\vec{K},\omega)$ has the form (in plasma resonance region).

$$\varepsilon(\vec{k},\omega)=\varepsilon_{\infty}(\vec{K},\omega)-\frac{\omega_p^2}{\omega^2}+i\varepsilon_2(\vec{K},\omega) \tag{4}$$

Our purpose is to determine the temperature dependence of -Im$\varepsilon^{-1}$ near the ferroelectric PT point when electron energy losses are determined by the plasmon excitation of valance electrons. $\varepsilon_\infty$ is attributed to virtual transitions from lower valence bands to the conduction band and depends on polarization:

$$\varepsilon_\infty=\varepsilon_\infty^p+\alpha P_s^2 \tag{5}$$

$$P_s=P_s(0)\tau \tag{6}$$

Where $P_s(0)$ is a polarization in a deep ferroelectric phase,

$\tau=(T_c-T)/T_c$

$T_c$ – is phase transition temperature. Therefore,

$$\varepsilon_\infty(\tau)=\varepsilon_\infty^p+(\varepsilon_\infty^f-\varepsilon_\infty^p)\tau \tag{7}$$

Substituting (7) into (1) we have

$$W=\frac{8\pi\pi^2}{\hbar K^2\varepsilon_2}\frac{1}{\left[\frac{\varepsilon_\infty^p+(\varepsilon_\infty^f-\varepsilon_\infty^p)\tau-(\varepsilon_p^2/\varepsilon^2)}{\varepsilon_2}\right]^2+1} \tag{8}$$

A linear size of the region where the energy transfer takes place, $\hbar\omega_p/\sqrt{\varepsilon_\infty}\sim 10\,\text{eV}$, has been taken into account.

Once can show that dependence $\varepsilon(\vec{K},\omega)$ on $\vec{K}$ is not considerable. Note that

$$[(\varepsilon_\infty^f-\varepsilon_\infty^p)/\varepsilon_\infty^p]<<1 \tag{9}$$

then, we can assert that $d\omega_s(\tau)/d\tau<<\omega_s(\tau)$, and $\omega_s^2(\tau)=\omega_p^2/\varepsilon_\infty(\tau)$.

Therefore, there exists such a frequency region $\Delta\omega$ when

$$|\omega_s(0)-\omega_s(1)|<<\Delta\omega<<\omega \tag{10}$$

As a frequency dependence of $\varepsilon_2(\omega)$ in the plasma resonance region is taken to scale of the frequency itself, then we can consider $\varepsilon_2$ independent of the frequency in the above range of $\Delta\omega$, and analyze the dependence of a real part of $\varepsilon(\omega)$ on $\omega_s(\tau)$.

Now let $\omega$ belong to the range $\Delta\omega$ (with all $\omega_s(\tau)$). Then, the expression (8) presents the dependence of W on $\tau$ at the above frequencies. From (10) we have

$$dW/d\tau|_{\tau=\tau_{max}}=0 \qquad d^2W/d\tau^2|_{\tau=\tau_{max}}<0 \tag{11}$$

at

$$\tau_{max}=[(\omega_p/\varepsilon)^2-\varepsilon_\infty^p]/(\varepsilon_\infty^f-\varepsilon_\infty^p) \tag{12}$$

Taking into account that $0<\tau_{max}<1$, we obtain that in the dependence $W(\tau)$ and the maximum is observed only for the frequencies

$$[\omega_p^2/\varepsilon_\infty^f]<\varepsilon^2<[\varepsilon_p^2/\varepsilon_\infty^p] \tag{13}$$

(we consider that $\varepsilon_\infty^f>\varepsilon_\infty^p$). One can obtain from (6) that in the mentioned frequency range

$$\frac{W(0)-W(1)}{W(1)}\sim\frac{\varepsilon_\infty^f-\varepsilon_\infty^p}{\varepsilon_2}\sim 10^{-1} \tag{14}$$

As seen from (14) the characteristic energy loss function anomaly estimations obtained at phase transition describe the experimental results on characteristic losses better than in [9].

## 7. Conclusion

In conclusion, we have performed an ab initio study of the electronic structures of some $ABO_3$ ferroelectric. Within the DFT-LDA framework, we have found it necessary to include the semicore states in the calculations. We have then followed the effect on the electronic structure as the crystal field evolves from cubic to tetragonal structure, to rhombohedral. We have described a fingerprint in the electronic structure of cubic and rhombohedral $KNbO_3$. By using our earlier EELS experimental results on $ABO_3$ ferroelectrics and the theoretical investigation in the present paper, we find plasmon oscillation energy for the investigated compounds. The differences between the phases in the electronic structures are reflected in the line shape of the valence plasmons and the anomalous behavior of plasmons was exhibited by the function representing the EEL in the region of PT.

Table 1. Atomic position and lattice parameters of some $ABO_3$ ferroelectrics [16-18,44]

| Crystals | Symmetry Class | Lattice Parameters (Å) | Atoms,WPa | Position |
|---|---|---|---|---|
| $KNbO_3$ | $m\bar{3}m$ | a=b=c=4.0214 | K(1a) | (0.0, 0.0, 0.0) |
| | | | Nb(1b) | (0.5, 0.5, 0.5) |
| | | | O(3c) | (0.5, 0.5, 0.0) |
| $KNbO_3$ | 4mm | a=b=3.9970 | K(1a) | (0.0, 0.0, 0.023) |
| | | c=4.0630 | Nb(1b) | (0.5, 0.5, 0.5) |
| | | | O1(1b) | (0.5, 0.5, 0.04) |
| | | | O2(2c) | (0.5, 0.0, 0.542) |
| $KNbO_3$ | 3m | a=b=c=4.016 | K | (0.0112, 0.0112, 0.0112) |
| | | | Nb | (0.5, 0.5, 0.5) |
| | | | O1 | (0.5295, 0.5295, 0.0308) |
| | | | O2 | (0.5295, 0.0308, 0.5295) |
| $LiNbO_3$ | 3c | a=b=c=5.4944 | Li(2a) | (0.2829, 0.2829, 0.2829) |
| | | | Nb(2a) | (0.0, 0.0, 0.0) |
| | | | O(6b) | (0.1139, 0.3601, -0.2799) |
| $LiTaO_3$ | 3c | a=b=c=5.4740 | Li(2a) | (0.2790, 0.2790, 0.2790) |
| | | | Ta(2a) | (0.0, 0.0, 0.0) |
| | | | O(6b) | (0.1188 0.3622, -0.2749) |
| $KTaO_3$ | $m\bar{3}m$ | a=b=c=3.988 | K | (0.0, 0.0, 0.0) |
| | | | Ta | (0.5, 0.5, 0.5) |
| | | | O | (0.5, 0.5, 0.0) |
| $BaTiO_3$ | 4mm | a=b=3.9909 | Ba(1a) | (0.0, 0.0, 0.0) |
| | | c=4.0352 | Ti(1b) | (0.5, 0.5, 0.5224) |
| | | | O1(1b) | 0.5, 0.5, -0.0244 |
| | | | O2(2c) | (0.5, 0.0, 0.4895) |
| $SrTiO_3$ | $m\bar{3}m$ | a=b=c=3.905 | Sr(1a) | (0.0, 0.0, 0.0) |
| | | | Ti(1b) | (0.5, 0.5, 0.5) |
| | | | O(3c) | (0.5, 0.5, 0.0) |
| $SrTiO_3$ | mcm | a=b=5.5225 | Sr(4b) | (0.0, 0.5, 0.5) |
| | | c=7.810 | Ti(4c) | (0.0, 0.0, 0.0) |
| | | | O1(4a) | (0.0, 0.0, 0.25) |
| | | | O2(8h) | (0.4944, 0.9944, 0.0 ) |

**Table 1**

Table 2. Band gap calculation results of some $ABO_3$ crystals.

| Materials | Phase | Symmetry Class | $E_g$, eV (indirect) | | $E_g$, eV (direct) |
|---|---|---|---|---|---|
| | | | Theory | Exp. | Theory |
| $KNbO_3$ | Cubic | Pm3m | 1.538 ( R-Γ) | - | 2.513(Γ-Γ) |
| | Tetragonal | P4mm | 1.538 ( R-Γ) | 3.30[28,44] | 2.513(Γ-Γ) |
| | Rhombohedral | R3m | 1.538 (Z-Γ) | - | 2.503(Γ-Γ) |
| $LiNbO_3$ | Rhombohedral | R3c | 3.390(Z-Γ) | 3.63[39] | 3.540(Γ-Γ) |
| $LiTaO_3$ | Rhombohedral | R3c | 3.840(Z-Γ) | 3.93[39] | 4.110(Γ-Γ) |
| $KTaO_3$ | Cubic | Pm3m | 2.158( R-Γ) | 3.79[40,44] | 2.987(Γ-Γ) |
| $BaTiO_3$ | Cubic | Pm3m | 1.920( R-Γ) | 3.05[44] | 2.041(Γ-Γ) |
| | Teragonal | P4mm | 2.145( R-Γ) | 3.26[44] | 2.294(Γ-Γ) |
| $SrTiO_3$ | Cubic | Pm3m | 1.895(R-Γ) | 3.37[44] | 2.230(Γ-Γ) |
| | Teragonal | I4/mcm | 2.2117(M-Γ) | - | 2.361(Γ-Γ) |

**Table 2**

Table 3. Theoretical and experimental data of plasmon energies and energy features of $-\mathrm{Im}\varepsilon^{-1}$ for some $ABO_3$ ferroelectrics.

| | Exp | Exp | Theory | Exp | Theoretical |
|---|---|---|---|---|---|
| Material | $E_p$, eV | $E_s$, eV * | $E_p$, eV | $-\mathrm{Im}\varepsilon^{-1}$ | $-\mathrm{Im}\varepsilon^{-1}$ |
| $KNbO_3$/Cubic | - | - | 21.8 | - | 6.5, 15.0, 22.6, 28.1, 41.5 |
| $KNbO_3$/Tetragonal | 24.0 | 17.0 | 22.0 | 7.1, 14.0, 23.2, 29.5 | 6.4, 14.9, 22.5, 28.0, 41.0 |
| $KNbO_3$/Rhombohedral | - | - | 21.9 | - | 6.4, 14.9, 22.5, 28.1, 41.4 |
| $LiNbO_3$ | 25.5 | 18.2 | 23.2 | 7.5, 11.0, 13.0, 14.6, 16.3, 25.5 | 7.65, 8.2, 19.4, 21.0, 21.7, 22.3 |
| $LiTaO_3$ | 25.3 | 18.0 | 24.0 | 7.1, 11.5, 15.3, 21.2, 25,5 | 7.95, 16.2, 17.8, 19.3, 21.6, 22.7, 23.9, 24.6 |
| $KTaO_3$ | 20.3 | 15.0 | 20.4 | 7.9, 14.7, 19.8, 21.5, 24.3, 31.2 | 6.2, 8.3, 11.7, 15.1, 21.3, 22.1, 23.4 |
| $BaTiO_3$ | 24.3 | 16.8 | 23.1 | 7.2, 12.3, 14.2, 19.6, 23.1, 28.0 | 6.5, 13.7, 21.2, 29.4 |
| $SrTiO_3$ | 29.5 | 21.0 | 22.6 | 6.1, 7.5, 14.2, 23.7, 29.5 | 7.82, 9.35, 12.7, 14.1, 17.2, 19.3, 22.4 |

**Table 3**

Figure Captions

Figure 1. Electronic band structure of cubic $KNbO_3$.

Figure 2. Electronic band structure of tetrahedral $KNbO_3$.

Figure 3. Electronic band structure of rhombohedral $KNbO_3$.

Figure 4. Total and partial DOS for cubic $KNbO_3$.

Figure 5. Total and partial DOS for tetrahedral $KNbO_3$.

Figure 6. Total and partial DOS for rombohedral $KNbO_3$.

Figure 7. Electron energy loss function spectra of $ABO_3$ ferroelectrics: experimental data by using Synchrotron Radiation [28].

Figure 8. The calculated real ($\varepsilon_1$) and imaginary ($\varepsilon_2$) parts of the dielectric function of $KNbO_3$ in cubic (a), tetragonal (b) and rhombohedral (c) phases.

Figure 9. The theoretical electron energy-loss spectra of $KNbO_3$ in the cubic (a), tetragonal (b) and rhombohedral (c) phases.

Figure 10. Temperature dependence of electron energy-loss function for $KNbO_3$ (experiment).

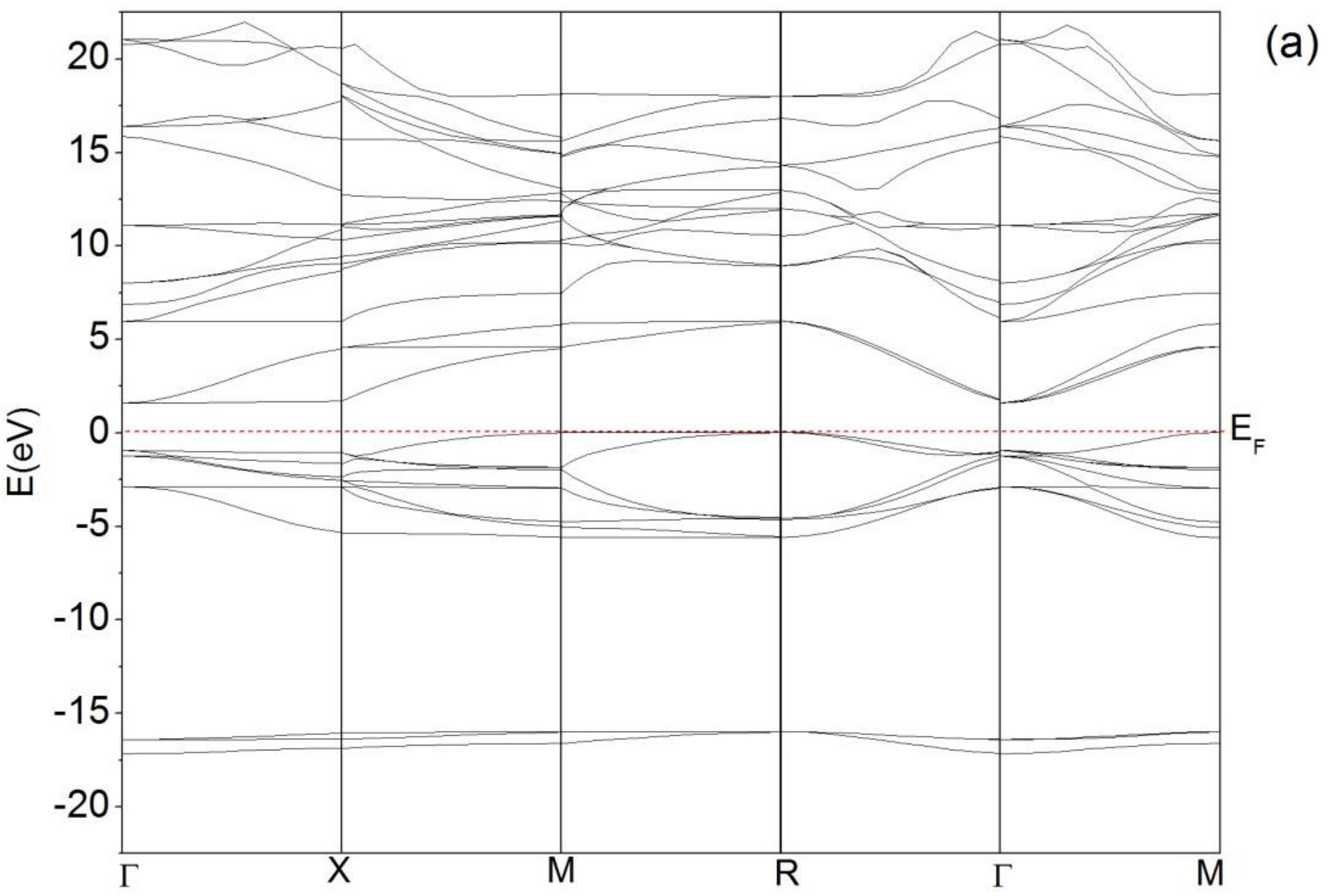

**Fİgure 1**

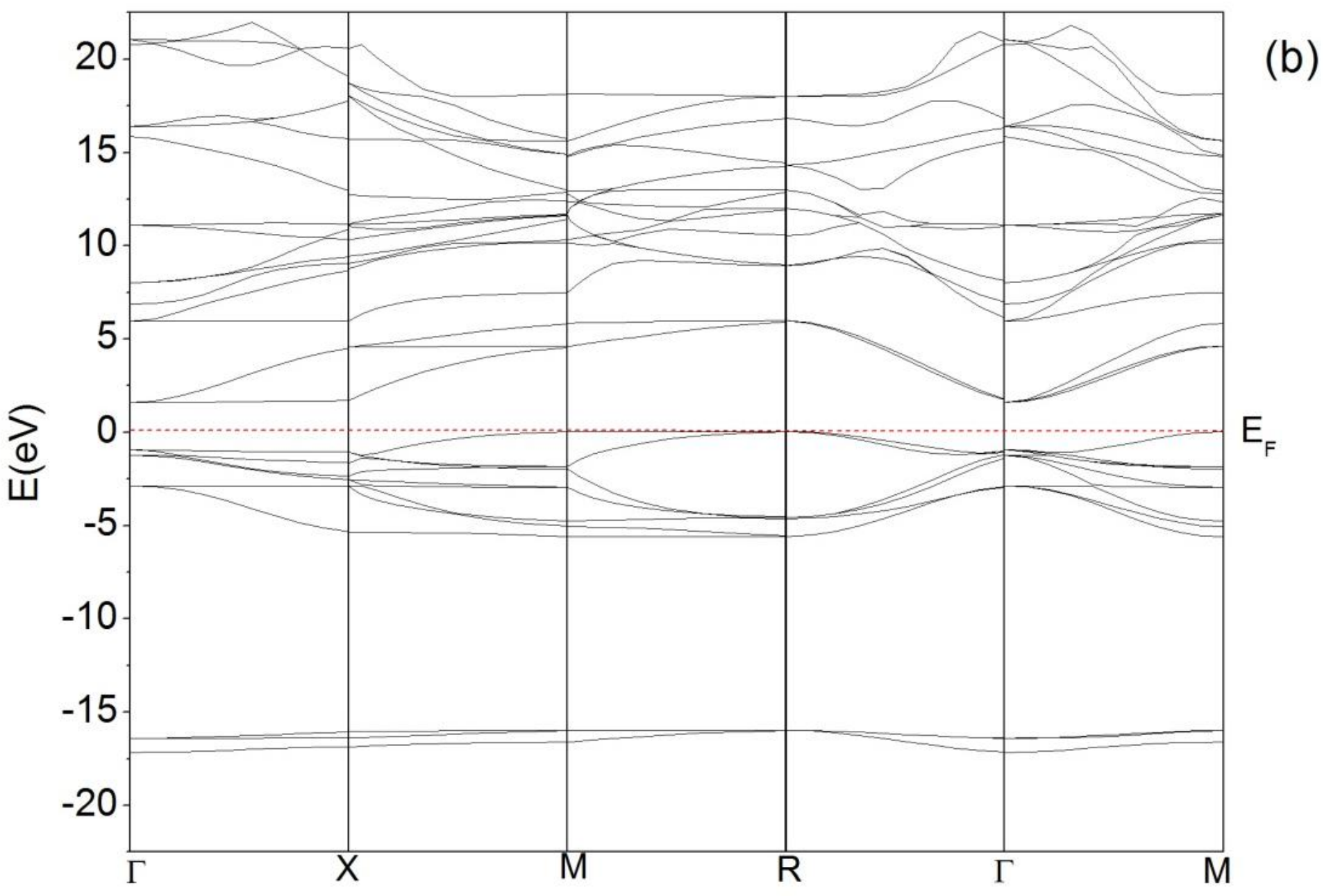

**Figure 2**

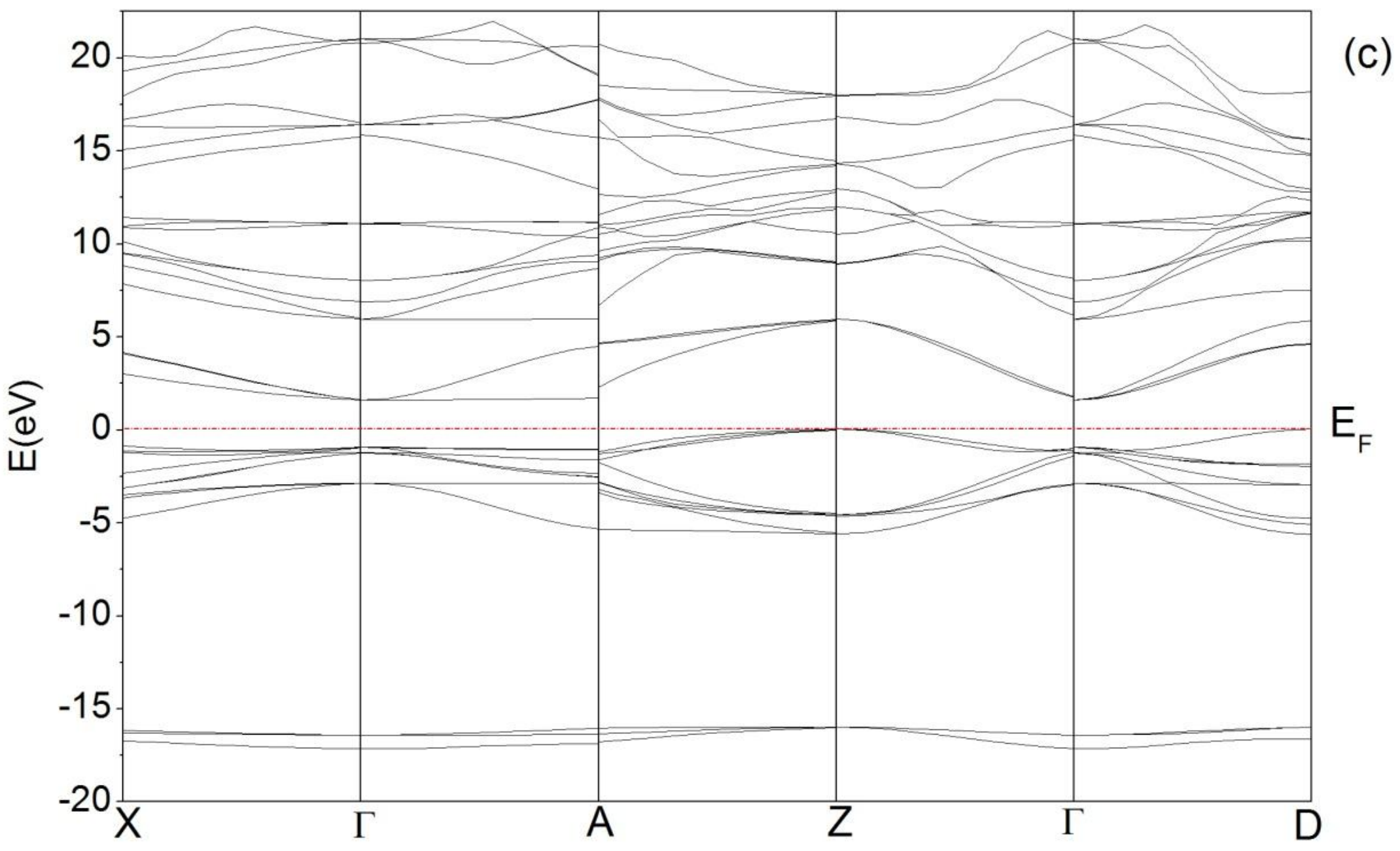

**Figure 3**

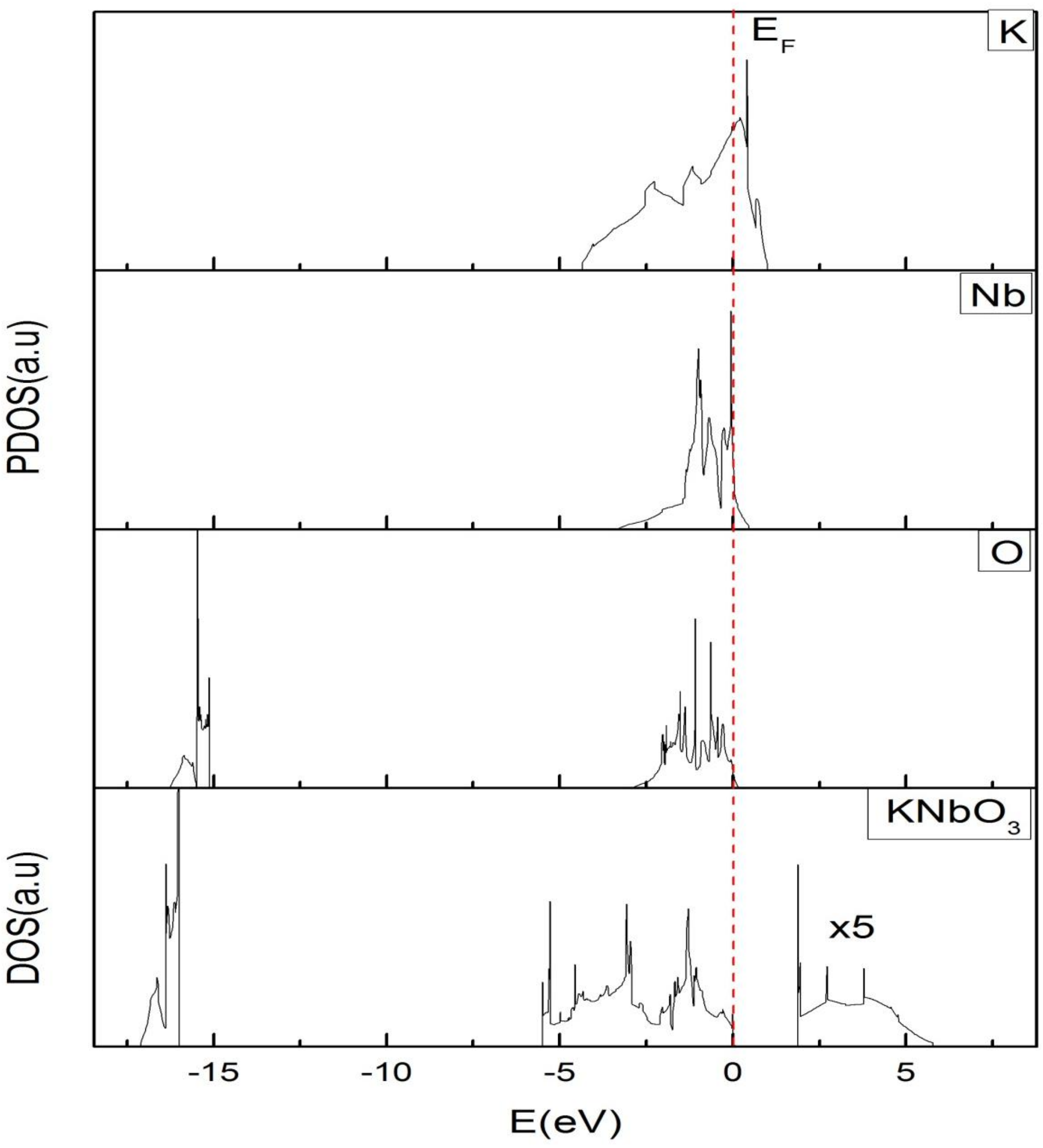

**Figure 4**

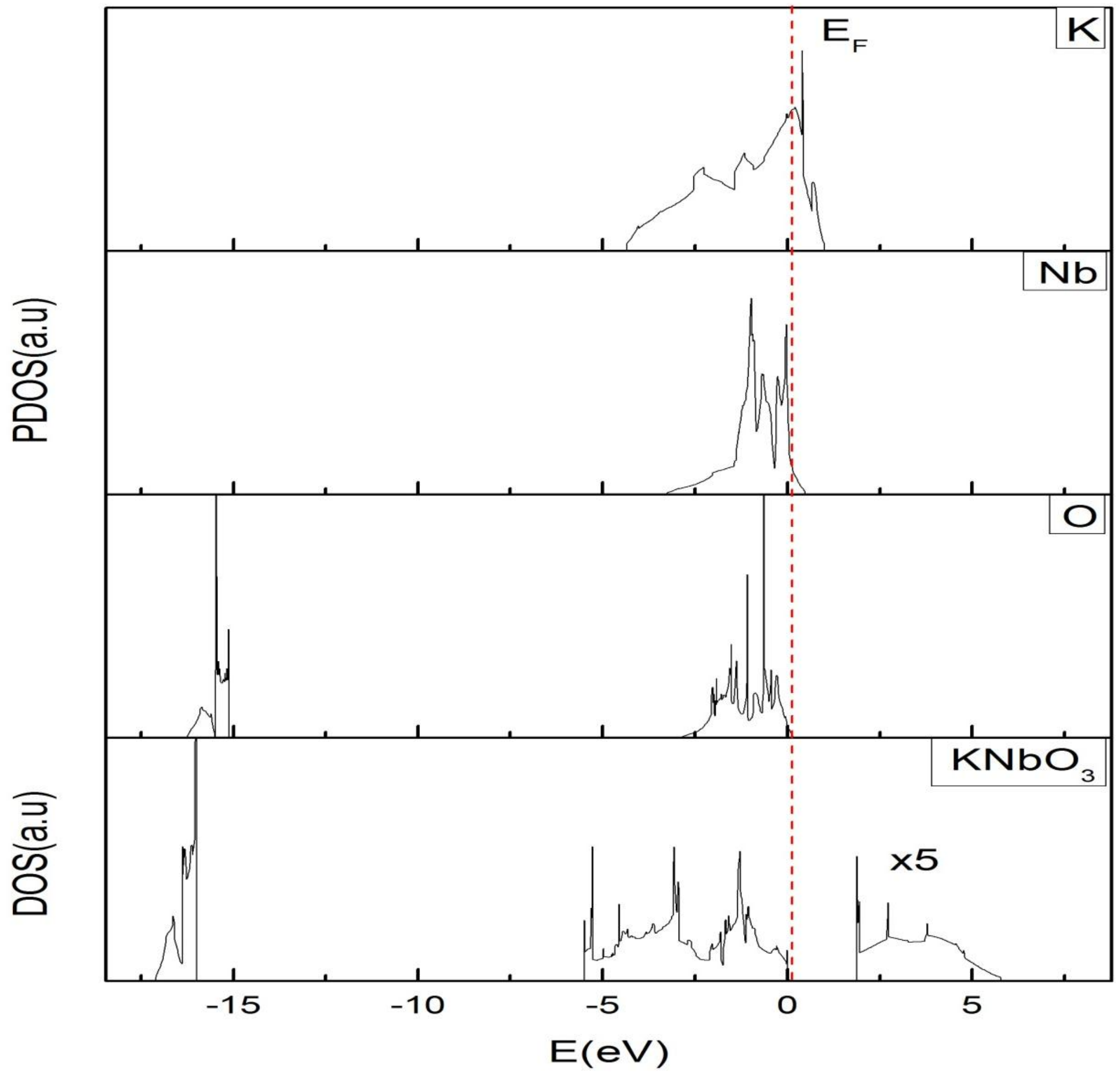

**Figure 5**

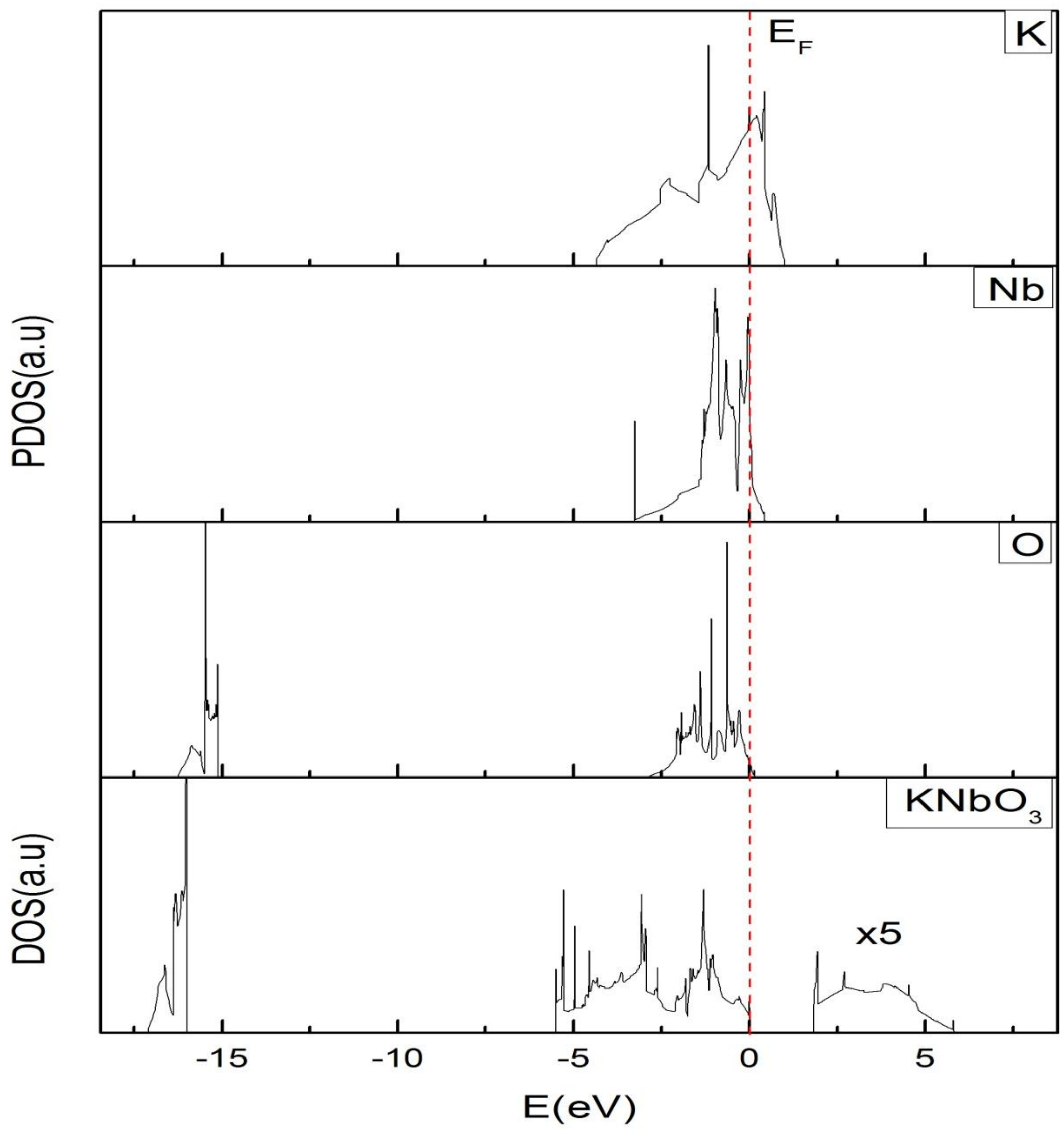

**Figure 6**

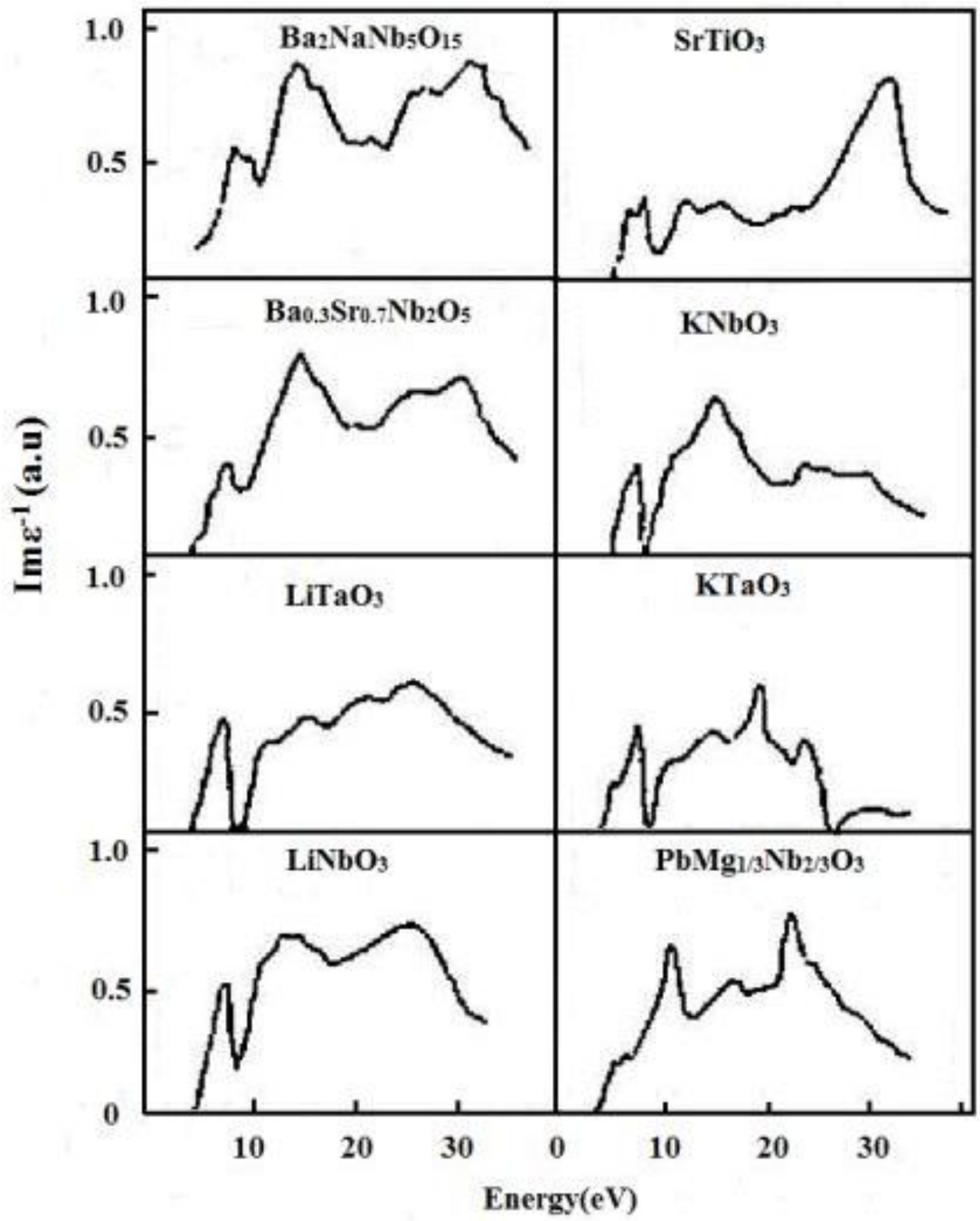

**Figure 7**

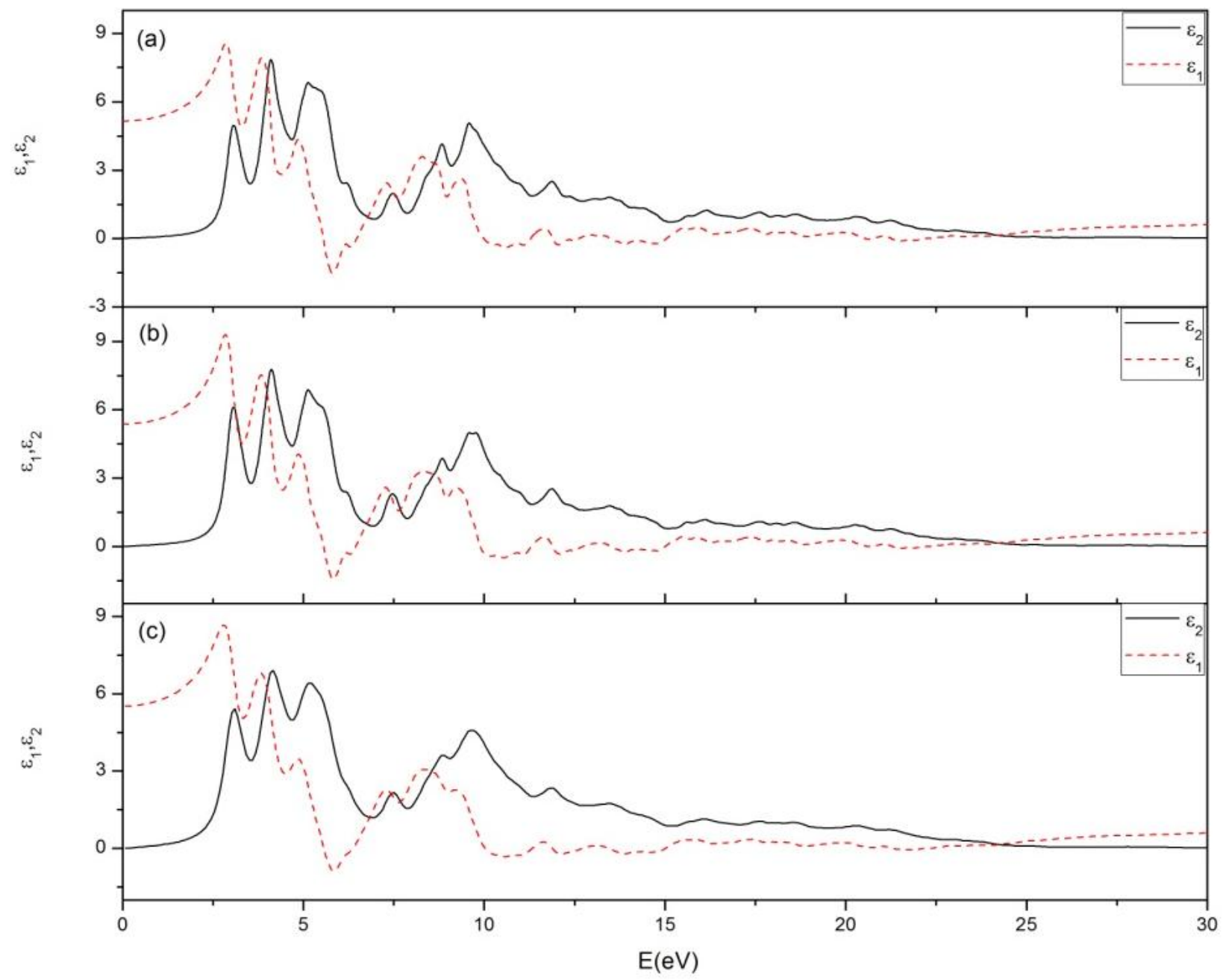

**Figure 8**

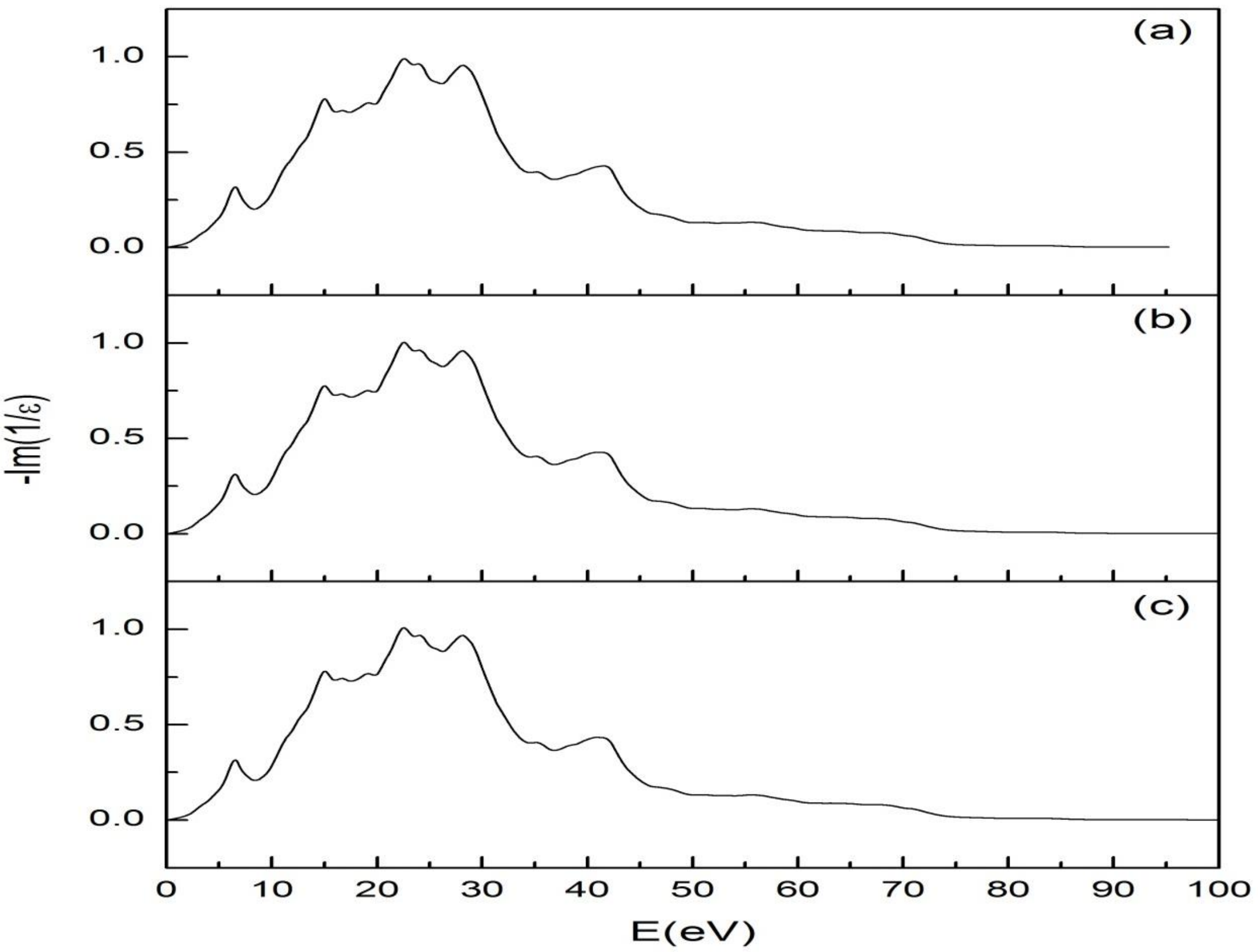

**Figure 9**

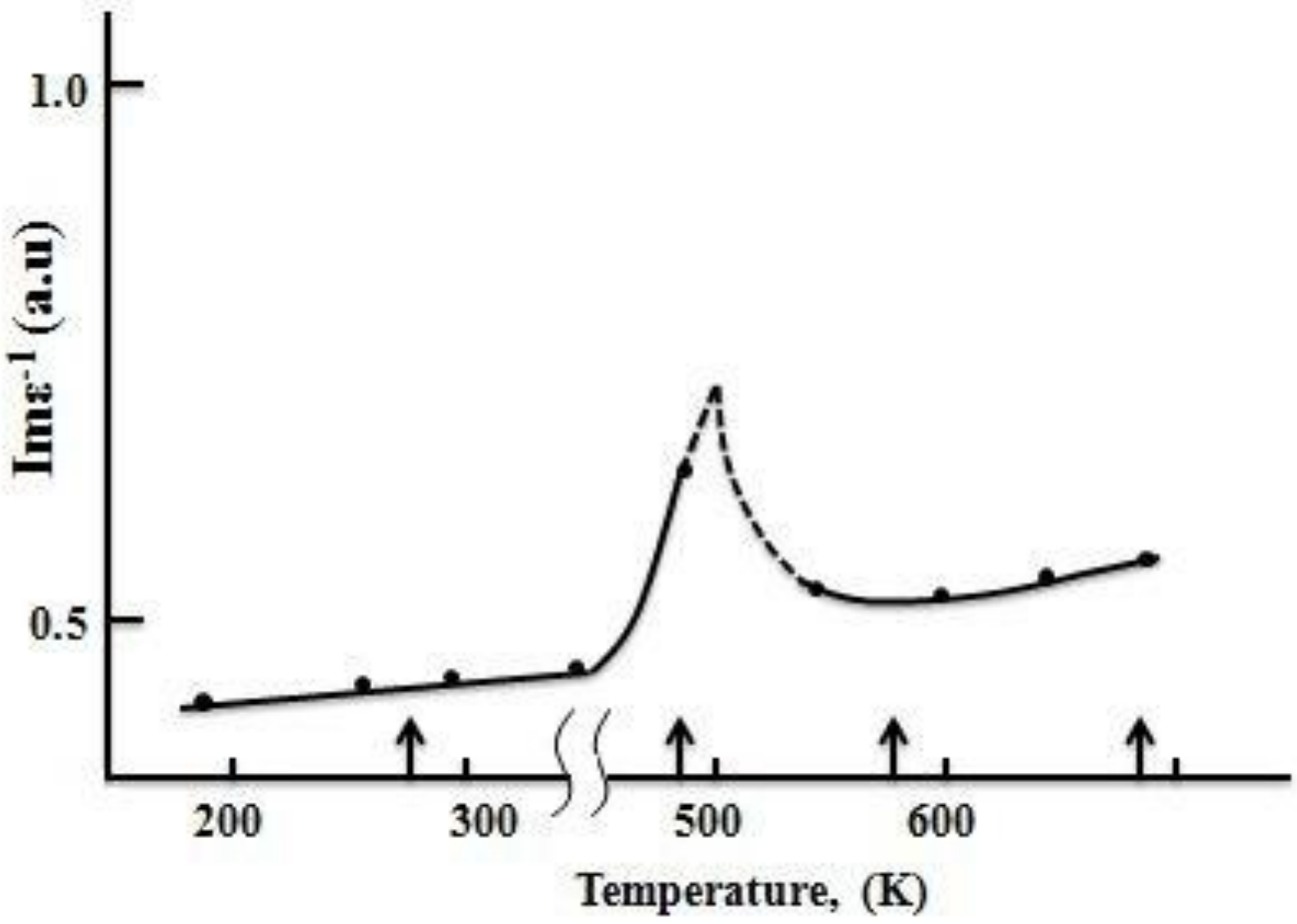

**Figure 10**